\documentclass[12pt]{iopart}
\usepackage{graphicx}
\begin{document}
\title[Self- and cross-phase modulation of ultrashort light pulses]
{Self- and cross-phase modulation of ultrashort light pulses in an inertial Kerr
medium:\\ spectral control of squeezed light}
\author{F Popescu and  A S Chirkin
\footnote[3]{Correspondence should addressed A.S.Chirkin}}
\address{Physics Department, M.V. Lomonosov Moscow State University,
Leninskie Gory, 119992 Moscow, Russia}
\eads{\mailto{florentin\_p@hotmail.com}, \mailto{chirkin@squeez.phys.msu.su}}

\begin{abstract}
Quantum theory of self-phase and cross-phase modulation of ultrashort light pulses in
the Kerr medium is developed with taking into account the response time of an
electronic nonlinearity. The correspondent algebra of time-dependent Bose-operators
is elaborated. It is established that the spectral region of the pulse, where the
quadrature fluctuations level is lower than the shot-noise one, depends on the value
of the nonlinear phase shift, the intensity of another pulse, and the relaxation time
of the nonlinearity. It is shown that the frequency of the pulse spectrum at which
the suppression of fluctuations is maximum can be controlled by adjusting the other
pulse intensity.
\end{abstract}


\submitto{\JOB}

{\sf Received: 23 October 2001, \quad Accepted: 19 March 2002}

\maketitle

\section{Introduction}

It is well known (see, for example, \cite{Tan,Kitagawa,ABC}) that in a nonlinear
medium with the Kerr nonlinearity, i.e. with cubic nonlinearity, self-phase
modulation (SPM) phenomenon occurs which gives rise to quadrature squeezing of
fluctuations with conservation of the photon statistics. This effect is observed when
the one-mode radiation propagates in the Kerr medium. If we deal with the propagation
of the two-mode radiation in such a medium, the situation becomes more complicated.
Besides the SPM effect, the phenomena of cross-phase modulation (XPM) and parametric
interactions can, generally speaking, take place. However, the efficiency of
parametric frequency-multiplication processes depends on the phase mismatch of the
interacting waves. The parametric frequency conversion can be neglected in the case
of the strong phase mismatch. It is worthy noting that the influence of the
parametric energy exchange at the two-mode interaction on the generation of
polarization-squeezed light in the Kerr medium has been considered in \cite{VolChir}.
Both SPM and XPM effects do not depend on the phase mismatch, consequently, they can
play a dominant role when one investigates the nonlinear propagation of two modes
with orthogonal polarizations and/or different frequencies in the Kerr medium. To the
best of our knowledge, such analysis has been carried out only for the case of
monochromatic modes (see review \cite{Tan2}).

In order to develop a correct quantum theory of the nonlinear propagation of the
pulse radiation, it is necessary to take into account a finite response time of the
nonlinearity. There are two consistent approaches to solving this problem. In
\cite{Boivin,Boivin1}\ the Kerr nonlinearity has been treated as a Raman -like
nonlinearity. The authors \cite{Boivin,Boivin1}\ have taken into account quantum and
thermal noises as a fluctuating addition to the relaxation nonlinearity in the
interaction Hamiltonian. In the other approach  developed in \cite{POP99,POP00,POP0},
the interaction Hamiltonian \cite{POP99}\ or the interaction momentum operator
\cite{POP00,POP0}, which contains only a response function of the nonlinearity, have
been used. In such quantum description of the SPM of ultra-short pulses (USPs), one
cannot add thermal noise terms to the interaction Hamiltonian and momentum operator
in order to satisfy the canonical commutation relation for the annihilation and
creation Bose-operators.

In this paper, we present the quantum theory based on the
interaction momentum operator \cite{POP00} for the case of
nonlinear propagation of two ultrashort light pulses
\cite{POP0}. In reality, here we will mainly analyze two
combined (SPM and XPM) effects. The first one is responsible for
the generation of the squeezed state of light and the second one
(as we demonstrate below) can be employed to control the
squeezing process. This combined phenomenon can be shortly
called ``the SPM-XPM of USPs.'' Since we deal with the analysis
of the combined phenomena, algebra of time dependent
Bose-operators developed earlier in \cite{POP00,Blow} has to be
subsequently extended.

The paper is organized as follows.

In section \ref{section2}\ the quantum equations for the SPM-XPM of USPs are
introduced when the nonlinearity of medium has relaxation behaviour of an electronic
Kerr type. Algebra of time-dependent Bose-operators is extended in section
\ref{section3} and used to estimate statistical parameters of the USP quadrature
components. In section \ref{section4} the correlation functions of the quadrature
components are obtained. Then in section \ref{section5} the spectrum of
quadrature-squeezed component of the USP under investigation is studied. Section
\ref{conclusions} closes the paper with some concluding remarks.

\section{Quantum equations for simultaneous SPM and XPM of
light pulses}\label{section2}

The conventional way to derive the quantum equation of SPM is based on the
interaction Hamiltonian, with the quantum equation describing the time evolution of
annihilation or creation Bose-operators being derived. A transition to the space
evolution is usually provided by the replacement $t\rightarrow z/u$, where $z$ is the
distance passed and $u$ is the pulse's group velocity in a nonlinear medium. This
approach seems to be fairly appropriate for the case of single-mode radiation.
Generally speaking, both $t$ and $z$ are present in the analytical description of
nonlinear propagation of the pulse. Therefore, the momentum operator connected with
the spatial evolution of the pulse field should be used instead of the interaction
Hamiltonian \cite{Mooki} in order to obtain an equation for the annihilation
Bose-operator. This approach for the case of pulse SPM in nonlinear medium with
inertial electronic Kerr nonlinearity has been successfully used in \cite{POP00}.

We start with the analysis of SPM--XPM of two coherent USPs in a noninertial
nonlinear medium.

\subsection{Quantum SPM-XPM equation in noninertial nonlinear medium}

In a noninertial nonlinear medium, the combined SPM--XPM effect of USPs is described
by making use of the momentum operator $\hat{G}(z)$ \cite{POP0}\
\begin{equation}\label{impuls}
\hat{G}(z)=\sum_{j=1}^{2}\hat{G}^{(j)}_{\rm spm}(z)+\hat{G}_{\rm xpm}(z),
\end{equation}
where index $j$ denotes the pulse's number ($j=1,2$), and the momentum operators of
SPM and XPM effects are
\begin{eqnarray}
\hat{G}^{(j)}_{\rm spm}(z)&=&\hbar\beta_{j}\int_{-\infty}^{\infty}
\hat{\mathbf{N}}[\hat{n}^{2}_{j}(t,z)]dt,\label{spm1}\\ \hat{G}_{\rm
xpm}(z)&=&\hbar\tilde{\beta}
\int_{-\infty}^{\infty}\hat{n}_{1}(t,z)\hat{n}_{2}(t,z)dt.\label{xpm1}
\end{eqnarray}
Here $\hbar$ is Planck's constant, $\hat{\mathbf{N}}$ is the operator of normal
ordering, the factors $\beta_{j}$ and $\tilde{\beta}$ are defined by the electronic
Kerr nonlinearity $n_{2}$ of the medium (for example,
$\beta_{j}=C^{2}\gamma^{\star}_{j}/2$; $C=(\hbar\omega_{0}/2V)^{1/2}$,
$\gamma^{\star}_{j}=k_0n_{2{(j)}}/2n_0$ \cite{POP00}), and
$\hat{n}_{j}(t,z)=\hat{A}^{+}_{j}(t,z)\hat{A}_{j}(t,z)$ is the photon number
``density" operator for the $j$-th pulse in a given cross-section $z$ at the time
moment $t$. The operator $\hat{A}^{+}_{j}(t,z)$ [$\hat{A}_{j}(t,z)$] is the photon
creation (annihilation) Bose operator. It is the slowly varying operator of the
negative (positive) frequency part of the electric field strength operator
\cite{Ahmanov}. The XPM momentum operator defined by (\ref{xpm1}) takes into account
the action of the first pulse on the second one and vice versa.

In the Heisenberg representation, the space evolution for the time-dependent
Bose-operator $\hat{A}_{j}(t,z)$ for the $j$-th pulse is given by the equation (see
\cite{Mooki})
\begin{equation}\label{interaction}
-i\hbar\frac{\partial\hat{A}_{j}(t,z)}{\partial
z}=\left[\hat{A}_{j}(t,z),\hat{G}(z)\right].
\end{equation}
In accordance with (\ref{impuls}), the space evolution equation, for example, of the
operator $\hat{A}_{1}(t,z)$ reads
\begin{equation}\label{eqs}
\frac{\partial\hat{A}_{1}(t,z)}{\partial
z}-\left[i\beta_{1}\hat{n}_{1}(t,z)
+i\tilde{\beta}\hat{n}_{2}(t,z)\right]\hat{A}_{1}(t,z)=0.
\end{equation}

The spatial evolution equation of the operator $\hat{A}_{2}(t,z)$ can be easily
obtained by permuting the indices $1\leftrightarrow 2$. Thus, we get the system of
four coupled equations keeping in mind two equations for the creation operator. One
points out that Eq.\ (\ref{eqs}) is written in the moving coordinate system:
$z=z^{'}$ and $t=t^{'}-z/u$, where $t^{'}$ is the running time, $u$ is the pulse's
group velocity in the nonlinear medium. We assume that one can neglect the difference
in group velocities of the pulses and do not consider the pulse spreading due to the
medium dispersion, that corresponds to first approximation of the dispersion theory.

In view of Eq.\ (\ref{eqs}), one can show that the operator $\hat{n}_{j}(t,z)$ does
not change in the nonlinear medium:
\begin{equation}\label{stat}
\hat{n}_{j}(t,z)=\hat{n}_{j}(t,z=0)=\hat{n}_{0,j}(t),
\end{equation}
where $z=0$ corresponds to the input of the nonlinear medium. Equation\ (\ref{stat})
means that the photon statistics does not change in the medium.

Solving Eq.\ (\ref{eqs}) and its Hermitian conjugate, one has for the annihilation
and creation photon Bose-operators
\begin{eqnarray}
\hat{A}_{1}(t,z)&=&e^{i\gamma_{1}\hat{n}_{0,1}(t)+
i\tilde{\gamma}\hat{n}_{0,2}(t)}
\,\hat{A}_{0,1}(t),\label{solution1}\\
\hat{A}^{+}_{1}(t,z)&=&\hat{A}^{+}_{0,1}(t)\,e^{-i\gamma_{1}
\hat{n}_{0,1}(t)-
i\tilde{\gamma}\hat{n}_{0,2}(t)},\label{solution2}
\end{eqnarray}
where $\gamma_{1}=\beta_{1}z$ and $\tilde{\gamma}=\tilde{\beta}z$. By permuting the
indices $1\leftrightarrow 2$, one gets the Bose-operators for another light pulse.

The preservation of the canonical structure of quantum theory requires that the
annihilation and creation Bose operators $\hat{A}_{j}(t,z)$ and
$\hat{A}^{+}_{j}(t,z)$ satisfy the commutation relations
\begin{equation}\label{delta}
\left[\hat{A}_{j}(t_{1},z),\hat{A}^{+}_{k}(t_{2},z)\right]=
\delta_{jk}\,\delta(t_{2}-t_{1})
\end{equation}
for an arbitrary distance $z$ in the medium, where $\delta_{jk}$ is the Kroneker
delta-symbol, $j,k=1,2$.

There are some peculiarities in the quantum description of the combined SPM--XPM
effect in a noninertial nonlinear medium. Firstly, the solutions (\ref{solution1}),
(\ref{solution2}) do not permit to verify the commutation relation (\ref{delta}).
Secondly, the reduction of the expressions $e^{i\gamma_{j}\hat{n}_{0,j}(t)}$,
$e^{i\tilde{\gamma}\hat{n}_{0,j}(t)}$ to the normally-ordered form is accompanied by
the appearance of a nonintegrable singularity (see \cite{POP00,Blow}). These
circumstances do not appear in the quantum theory of combined SPM-XPM effect which
takes into account the relaxation behaviour of the nonlinearity.

\subsection{Quantum equation of pulse SPM-XPM in inertial nonlinear
medium}

The momentum operator for SPM of USP in a medium with the electronic Kerr
nonlinearity has been introduced in \cite{POP00}. It incorporates the function of a
nonlinear response in its structure. In the model considered, a contribution of the
electronic Kerr nonlinearity decreases exponentially. Thus, we rewrite the momentum
operator of SPM--XPM effect $\hat{G}(z)$ introduced according to (\ref{impuls}) (see
\cite{POP0}) as follows
\begin{eqnarray}
\fl \hat{G}^{(j)}_{\rm{spm}}(z)=\hbar\beta_{j}\int_{-\infty}^{\infty}dt
\int_{-\infty}^{t}H(t-t_{1})\hat{\mathbf{N}}\bigl
[\hat{n}_{j}(t,z)\hat{n}_{j}(t_{1},z)\bigl]dt_{1},\label{spm2}\\ \fl
\hat{G}_{\rm{xpm}}(z)=\hbar\tilde{\beta}\int_{-\infty}^{\infty}dt
\int_{-\infty}^{t}H(t-t_{1})\bigl[\hat{n}_{1}(t,z)\hat{n}_{2}(t_{1},z)
+\hat{n}_{1}(t_{1},z)\hat{n}_{2}(t,z)\bigl]\,dt_{1},\label{xpm2}
\end{eqnarray}
where $H(t)$ is the function of nonlinear response, asymmetrically defined in order
to satisfy the condition imposed by the causality principle: $H(t)\neq 0$ at $t\geq
0$ and $H(t)=0$ at $t<0$. The term in the second integral in (\ref{spm2}) should be
interpreted as a generalized force acting in the cross-section $z$ of the medium
which at the time moment $t$ depends only on the previous time moments. The similar
term in the second integral in (\ref{xpm2}) should be interpreted as a sum of two
generalized forces acting between pulses in the same cross-section $z$ of a medium
which also depend on the previous time moments. Therefore, the causality principle in
the Hermitian operators (\ref{spm2}) and (\ref{xpm2}) is not violated.

In the case of the nonlinearity of an electronic origin, the nonlinear response
function can be introduced as follows:
\begin{equation}\label{has}
H(t)=(1/\tau_r)\exp{\left(-t/\tau_r\right)},\qquad(t\geq 0)
\end{equation}
and $H(t)=0$ at $t<0$. Indeed, if a single USP propagates through the Kerr medium,
then the evolution of nonlinear addition $\Delta n_{nl}$ to the refractive index,
which is associated with the SPM effect, is given by the equation (see, for example,
\cite{POP00,Blow,Ahmanov})
\begin{equation}\label{rel-eq}
\tau_{r}\frac{\partial\Delta\hat{n}_{nl}(t,z)}{\partial t}+\Delta
\hat{n}_{nl}(t,z)=\frac{1}{2}\,n_{2{(j)}}\,\hat{n}_{j}(t,z),
\end{equation}
which has the solution
\begin{equation}\label{add1}
\Delta\hat{n}_{nl}(t,z)=\frac{1}{2}\,n_{2{(j)}}\int_{-\infty}^{t}
H(t-t_{1})\hat{n}_{j}(t_{1},z)dt_{1}.
\end{equation}

Note that the nonlinear response function (\ref{has}) appears in the absence of one-
and two-photon and Raman resonances ~\cite{Ahmanov}. Therefore, the approach
developed is valid when each pulse frequency is off-resonance and the pulse duration
$\tau_{p}$ is much larger than the relaxation time $\tau_{r}$. If the USP propagates
in a fused silica-fiber, then about $80$\% of the Kerr effect is due to the
electronic motion occurring on $\sim 1$ fs time scales and only $20$\% of the Kerr
effect is attributable to the Raman oscillators \cite{Joneckis}. Thus, our model
corresponds to the case of the Kerr effect mainly produced by the electronic motion.

Since in quantum theory the relaxation behaviour is connected with the so-called
thermal reservoir, the operator evolution equation (\ref{rel-eq}) must, in general
case, contain a source of thermal noise besides a relaxation term. Then we can write
the following expression for the nonlinear addition \cite{Boivin,Boivin1}
\begin{equation}\label{add2}
\Delta\hat{n}_{nl}(t,z)=\frac{1}{2}\,n_{2{(j)}}\int_{-\infty}^{t}
H(t-t_{1})\hat{n}_{j}(t_{1},z)dt_{1}\,+\,\hat{m}(t,z),
\end{equation}
where the Hermitian operator $\hat{m}(t,z)$ takes into account thermal fluctuations
of $\Delta\hat{n}_{nl}(t,z)$ in the absence of light field. For the nonlinearity of
electronic origin an expression likes (\ref{add2}) can be obtained from Duffing-type
equation \cite{Ahmanov}. Here it is important to note that the average value of
$\hat{m}(t,z)$ operator is equal to zero; $\langle\hat{m}(t,z)\rangle=0$ (see also
\cite{Boivin1}). Hence, one can consider Eq.\ (\ref{add1}) as the result of averaging
Eq.\ (\ref{add2}) over the thermal fluctuations. In connection with that, expressions
(\ref{spm2}), (\ref{xpm2}) can be truly considered as ones averaged over the thermal
fluctuations of
 nonlinearity.

In the present paper  we neglect the thermal fluctuations since we are interested in
the nonlinear phase fluctuations caused by the quantum USP ones. Nevertheless, the
commutation relations (\ref{delta}) are proved to be valued in this simplified
approach (see below). However, it is not difficult to generalize the theory of the
SPM and XPM phenomena allowing for thermal noise in the Kerr medium with electronic
nonlinearity.

Using (\ref{delta}) it is easy to prove that the operator $\hat{n}_{j}(t,z)$ commutes
with the the momentum operator $\hat{G}(z)$,
\begin{equation}\label{opera}
[\hat{G}(z),\hat{n}_{j}(t,z)]=[\hat{G}(z),\hat{n}_{0,j}(t)]=0,
\end{equation}
that is, the photon number operator $\hat{n}_{j}(t,z)$ remains unchanged in the
nonlinear medium [see also Eq.\ (\ref{stat})].

Taking into account (\ref{spm2}) and (\ref{xpm2}), one obtains from
(\ref{interaction}) the quantum equation for combined SPM--XPM effect, for instance,
for the pulse with index $1$
\begin{equation}\label{part1}
\frac{\partial\hat{A}_{1}(t,z)}{\partial z}-\left\{i\beta_{1}q[\hat{n}_{0,1}(t)]
+i\tilde{\beta}q[\hat{n}_{0,2}(t)]\right\}\hat{A}_{1}(t,z)=0,
\end{equation}
where
\begin{equation}\label{quext}
q[\hat{n}_{0,j}(t)]=\int_{0}^{\infty}H(t_{1})\left[\hat{n}_{0,j}(t-t_{1})+
\hat{n}_{0,j}(t+t_{1})\right]dt_{1}.
\end{equation}
The appearance of the second term in (\ref{quext}) related to
$\hat{n}_{0,j}(t+t_{1})$ in the quantum description has been already  discussed in
\cite{POP00} where it has been assumed that this term can be connected with the
vacuum fluctuations which are present even in the absence of a pulse. It should be
also pointed out that expression (\ref{part1}) is an intermediate result written in
the moving coordinate system.

Solving the space evolution equation (\ref{part1}) for the annihilation Bose-operator
and its Hermitian conjugate we obtain
\begin{eqnarray}
\hat{A}_{1}(t,z)&=&e^{i\gamma_{1}q[\hat{n}_{0,1}(t)]+
i\tilde{\gamma}q[\hat{n}_{0,2}(t)]}\,\hat{A}_{0,1}(t),\label{winzip1}\\
\hat{A}^{+}_{1}(t,z)&=&\hat{A}^{+}_{0,1}(t)\,e^{-i\gamma_{1}q[\hat{n}_{0,1}(t)]-
i\tilde{\gamma}q[\hat{n}_{0,2}(t)]}.\label{winzip2}
\end{eqnarray}
One can obtain similar expressions for the Bose-operators of the second pulse by
permuting indices $1\leftrightarrow 2$. It is convenient to rewrite expression
(\ref{quext}) as follows
\begin{equation}\label{qup}
q[\hat{n}_{0,j}(t)]=\int_{-\infty}^{\infty}h(t_{1})\hat{n}_{0,j}(t-t_{1})dt_{1},
\qquad [h(t)=H(\left|t\right|)].
\end{equation}
If we consider $\hat{n}_{0,j}$ to be time independent, then (\ref{winzip1}) and
(\ref{winzip2}) describe the case of monochromatic modes (see, for example,
\cite{Tan2,Orlov}). If in (\ref{winzip1}) and (\ref{winzip2}) the response function
$H(t)=\delta(t)$, then the description of the noninertial nonlinear media
(\ref{solution1}) and (\ref{solution2}) can be produced. Note that in the quantum
description the structure of the nonlinear response (\ref{qup}) is similar to the one
of a linear response of a medium in second-order approximation of the dispersion
theory (see \cite{POP00,Ahmanov}).

To estimate the statistical characteristics of pulses at the output of the nonlinear
medium, we need to calculate the average values of the operators' moments. As it is
well known, they can be found if the operator expressions are given in the normally
ordered form. The use of solutions (\ref{winzip1}) and (\ref{winzip2}) involves the
development of a special mathematical technique. Bellow some elements of algebra of
time-dependent Bose-operators developed in \cite{POP00} are extended.

\section{Algebra of time-dependent Bose-operators}\label{section3}

For the quantum analysis of the pulse's SPM the correspondent algebra of
time-dependent Bose-operators has been developed in \cite{POP00}. In the case where
the XPM effect is present in addition to the SPM effect, we need to improve the
mentioned algebra. It is convenient to introduce the operators
\begin{equation}
\hat{O}_{j}(t)=i\gamma_{j}\,q[\hat{n}_{0,j}(t)],\qquad
\hat{\tilde{O}}_{j}(t)=i\tilde{\gamma}q[\hat{n}_{0,j}(t)],\label{ou1}
\end{equation}
and their Hermitian conjugates
\begin{equation}
\hat{O}^{+}_{j}(t)=-\hat{O}_{j}(t),\qquad
\hat{\tilde{O}}^{+}_{j}(t)=-\hat{\tilde{O}}_{j}(t),\label{ou3}
\end{equation}
in order to simplify further calculations. Thus, Eqs.\ (\ref{winzip1}),
(\ref{winzip2}) may be represented in the operator form:
\begin{eqnarray}
\hat{A}_{1}(t,z)&=&e^{\hat{O}_{1}(t)+
\hat{\tilde{O}}_{2}(t)}\,\hat{A}_{0,1}(t),\label{pulse1a}\\
\hat{A}^{+}_{1}(t,z)&=&\hat{A}^{+}_{0,1}(t)\,e^{\hat{O}^{+}_{1}(t)+
\hat{\tilde{O}}^{+}_{2}(t)}\label{pulse1b},
\end{eqnarray}
where the operator $\hat{O}_{1}(t)$ is responsible for the SPM of the $j$-th pulse
and the operator $\hat{\tilde{O}}_{2}(t)$ corresponds to the XPM effect due to the
action of the second (control) pulse on the first  pulse (under investigation).

We suppose that the initial pulses are in coherent states, the $j$-th pulse's
operator $\hat{A}_{0,j}(t)$ acting only on the state vector $|\alpha_{0,j}(t)\rangle$
within the associated Hilbert space ${\mathcal{H}_{j}}$:
\[\hat{A}_{0,j}(t)|\alpha_{0,j}(t)\rangle=
\alpha_{0,j}(t)|\alpha_{0,j}(t)\rangle,\] where  $\alpha_{0,j}(t)$ is the eigenvalue
of the operator. The factorization of the states of two different sub-Hilbert spaces
${\mathcal{H}_{1}}$ and ${\mathcal{H}_{2}}$ takes place. The summarized quantum state
of two pulses is described by the vector
\[|\alpha_{0}(t)\rangle=|\alpha_{0,1}(t)\rangle \otimes|\alpha_{0,2}(t)\rangle\] in
the global Hilbert space ${\mathcal{H}}={\mathcal{H}_{1}}\otimes{\mathcal{H}_{2}}$.
Since the coherent quantum states of each pulse occupy a distinct sub-Hilbert space
for any two arbitrary chosen time moments $t_{1}$ and $t_{2}$ one can write :
\begin{equation}
[\hat{n}_{1}(t_{1},z),\hat{n}_{2}(t_{2},z)]=[\hat{n}_{0,1}(t_{1}),\hat{n}_{0,2}(t_{2})]=0,
\end{equation}
from which we obtain
\begin{eqnarray}
\bigl[\hat{A}_{0,1}(t_{1}),e^{\hat{O}_{2}(t_{2})}\bigl]=0,&\qquad&
\bigl[\hat{A}_{0,2}(t_{1}),e^{\hat{O}_{1}(t_{2})}\bigl]=0,\label{co1}\\
\bigl[\hat{A}^{+}_{0,1}(t_{1}),e^{\hat{O}_{2}(t_{2})}\bigl]=0,&\qquad&
\bigl[\hat{A}^{+}_{0,2}(t_{1}),e^{\hat{O}_{1}(t_{2})}\bigl]=0.\label{co2}
\end{eqnarray}
Calculating (\ref{co1}) and (\ref{co2}) it is easy to prove that the following
relationships take place:
\begin{eqnarray}
\bigl[e^{\hat{O}_{1}(t_{1})},e^{\hat{O}_{2}(t_{2})}\bigl]&=&
\bigl[e^{\hat{O}^{+}_{1}(t_{1})},e^{\hat{O}^{+}_{2}(t_{2})}\bigl]=0,\label{con}\\
\bigl[e^{\hat{\tilde{O}}_{1}(t_{1})},e^{\hat{\tilde{O}}_{2}(t_{2})}\bigl]
&=&\bigl[e^{\hat{\tilde{O}}^{+}_{1}(t_{1})},e^{\hat{\tilde{O}}^{+}_{2}(t_{2})}\bigl]=0.\label{com}
\end{eqnarray}
It should be pointed out that in the next sections the averaging operations are made
over the total quantum state $|\alpha_{0}(t)\rangle$.

\subsection{Operator permutation relations}

For each pulse, the following operator permutation relations hold (see also
\cite{POP00}):
\begin{eqnarray}
\hat{A}_{0,j}(t_{1})\hat{O}_{j}(t_{2})&=&[\hat{O}_{j}(t_{2})+{\mathcal{D}}_{j}
(t_{2}-t_{1})]\hat{A}_{0,j}(t_{1}),\\
\hat{O}_{j}(t_{1})\hat{A}^{+}_{0,j}(t_{2})&=&\hat{A}^{+}_{0,j}(t_{2})
[\hat{O}_{j}(t_{1})+{\mathcal{D}}_{j}(t_{2}-t_{1})],
\end{eqnarray}
where \[{\mathcal{D}}_{j}(t_{2}-t_{1})=i\gamma_{j}\,h(t_{2}-t_{1}).\] Besides,
$\mathcal{D}_{j}(t_{2}-t_{1})=\mathcal{D}_{j}(t_{1}-t_{2})$ since $h(t)$ is an even
function of $t$ [see (\ref{qup})]. In view of the mathematical induction, it is
possible to demonstrate the validity of the formulae ($m\in{\mathrm{N}}$):
\begin{eqnarray}
\hat{A}_{0,j}(t_{1})\hat{O}^{m}_{j}(t_{2})&=&[\hat{O}_{j}(t_{2})+
{\mathcal{D}}_{j}(t_{2}-t_{1})]^{m}\hat{A}_{0,j}(t_{1}),\label{advance1}\\
\hat{O}^{m}_{j}(t_{1})\hat{A}^{+}_{0,j}(t_{2})&=&\hat{A}^{+}_{0,j}(t_{2})
[\hat{O}_{j}(t_{1})+{\mathcal{D}}_{j}(t_{2}-t_{1})]^{m}.\label{advance2}
\end{eqnarray}
Expanding $e^{\hat{O}_{j}(t)}$ and $e^{\hat{O}^{+}_{j}(t)}$ in Taylor series, we can
get the operator permutation relations which play an important role while estimating
the statistical characteristics of the pulse investigated. By using (\ref{advance1})
and (\ref{advance2}) we finally arrive at
\begin{eqnarray}
\hat{A}_{0,j}(t_{1})e^{\hat{O}_{j}(t_{2})}&=&e^{\hat{O}_{j}(t_{2})
+{\mathcal{D}}_{j}(t_{2}-t_{1})}\hat{A}_{0,j}(t_{1}),\label{permut1}\\
e^{\hat{O}_{j}(t_{1})}\hat{A}^{+}_{0,j}(t_{2})&=&\hat{A}^{+}_{0,j}(t_{2})
e^{\hat{O}_{j}(t_{1})+{\mathcal{D}}_{j}(t_{2}-t_{1})}.\label{permut2}
\end{eqnarray}
Other permutation relations can be obtained by Hermitian conjugation. Making use of
the relations (\ref{permut1}), (\ref{permut2}) and Eqs.\ (\ref{con}), (\ref{com}) one
can show that the canonical commutation relation (\ref{delta}) for the operators
$\hat{A}_{j}(t,z)$ and $\hat{A}^{+}_{j}(t,z)$ is exactly fulfilled. Proceeding in the
same manner we can prove the validity of the following relations
\begin{eqnarray}\label{addi}
\bigl[\hat{A}_{j}(t_{1},z),\hat{A}_{j}(t_{2},z)\bigl]&=&0,\label{nucu1}\\ {}
\bigl[\hat{n}_{j}(t_{1},z),\,\hat{n}_{j}(t_{2},z)\bigl]&=&0.\label{nucu2}
\end{eqnarray}
Let j=1. Then
\begin{eqnarray}
\fl\bigl[\hat{A}_{1}(t_{1},z),\hat{A}_{1}(t_{2},z)\bigl]&=&\hat{A}_{1}(t_{1},z)\hat{A}_{1}(t_{2},z)
-\hat{A}_{1}(t_{2},z)\hat{A}_{1}(t_{1},z)\nonumber\\
\fl&=&e^{\hat{O}_{1}(t_{1})+\hat{\tilde{O}}_{2}(t_{1})}\hat{A}_{0,1}(t_{1})e^{\hat{O}_{1}(t_{2})+
\hat{\tilde{O}}_{2}(t_{2})}\hat{A}_{0,1}(t_{2})\nonumber\\
\fl&{}&-e^{\hat{O}_{1}(t_{2})+
\hat{\tilde{O}}_{2}(t_{2})}\hat{A}_{0,1}(t_{2})e^{\hat{O}_{1}(t_{1})+
\hat{\tilde{O}}_{2}(t_{1})}\hat{A}_{0,1}(t_{1})\nonumber\\
\fl&=&e^{\hat{O}_{1}(t_{1})+\hat{O}_{1}(t_{2})+\hat{\tilde{O}}_{2}(t_{1})+
\hat{\tilde{O}}_{2}(t_{2})}e^{\mathcal{D}_{1}(t_{2}-t_{1})}\hat{A}_{0,1}(t_{1})\hat{A}_{0,1}(t_{2})\nonumber\\
&{}&-e^{\hat{O}_{1}(t_{1})+\hat{O}_{1}(t_{2})+\hat{\tilde{O}}_{2}(t_{1})+
\hat{\tilde{O}}_{2}(t_{2})}e^{\mathcal{D}_{1}(t_{1}-t_{2})}\hat{A}_{0,1}(t_{1})\hat{A}_{0,1}(t_{2})\nonumber\\
&=& 0.
\end{eqnarray}
We used above the permutation relation (\ref{permut1}). Let us verify our initial
statement (\ref{stat}).
\begin{eqnarray}\label{hy}
\hat{n}_{1}(t,z)&=&\hat{A}^{+}_{1}(t,z)\hat{A}_{1}(t,z)\nonumber\\
&=&\hat{A}^{+}_{0,1}(t)e^{\hat{O}^{+}_{1}(t)+
\hat{\tilde{O}}^{+}_{2}(t)}e^{\hat{O}_{1}(t)+
\hat{\tilde{O}}_{2}(t)}\hat{A}_{0,1}(t)\nonumber\\
&=&\hat{A}^{+}_{0,1}(t)e^{\hat{O}_{1}(t)+\hat{O}^{+}_{1}(t)+
\hat{\tilde{O}}_{2}(t)+\hat{\tilde{O}}^{+}_{2}(t)}\hat{A}_{0,1}(t)\nonumber\\
&=&\hat{A}^{+}_{0,1}(t)\hat{A}_{0,1}(t)\nonumber\\ &=&\hat{n}_{0,1}(t).
\end{eqnarray}
Hence we have
\begin{eqnarray}\label{vy}
\fl\bigl[\hat{n}_{1}(t_{1},z),\hat{n}_{1}(t_{2},z)\bigl]&=&[\hat{n}_{0,1}(t_{1}),\hat{n}_{0,1}(t_{2})]\nonumber\\
&=&\hat{n}_{0,1}(t_{1})\hat{n}_{0,1}(t_{2})
-\hat{n}_{0,1}(t_{2})\hat{n}_{0,1}(t_{1})\nonumber\\
\fl&=&\hat{A}^{+}_{0,1}(t_{1})\hat{A}_{0,1}(t_{1})\hat{A}_{0,1}^{+}(t_{2})\hat{A}_{0,1}(t_{2})\nonumber\\
\fl&{}&-\hat{A}_{0,1}^{+}(t_{2})\hat{A}_{0,1}(t_{2})\hat{A}^{+}_{0,1}(t_{1})\hat{A}_{0,1}(t_{1})\nonumber\\
\fl&=&\hat{A}^{+}_{0,1}(t_{1})\hat{A}_{0,1}^{+}(t_{2})\hat{A}_{0,1}(t_{1})\hat{A}_{0,1}(t_{2})
+\delta(t_{2}-t_{1})\hat{A}^{+}_{0,1}(t_{1})\hat{A}_{0,1}(t_{2})\nonumber\\
&{}&-\hat{A}^{+}_{0,1}(t_{1})\hat{A}_{0,1}^{+}(t_{2})\hat{A}_{0,1}(t_{1})\hat{A}_{0,1}(t_{2})-
\delta(t_{1}-t_{2})\hat{A}^{+}_{0,1}(t_{2})\hat{A}_{0,1}(t_{1})\nonumber\\ &=&0.
\end{eqnarray}
In (\ref{vy}) we used the fact that for any functions $f(x)$, $g(x)$ we have
$\delta(x_{2}-x_{1})f(x_{1})g(x_{2})=\delta(x_{1}-x_{2})f(x_{2})g(x_{1})$, where
$x_{1}$ and $x_{2}$ are two arbitrary values in which the functions are defined.

\subsection{Normal ordering}

As it was stated above, another important issue is presenting the operators
$\hat{A}_{j}(t,z)$ and $\hat{A}^{+}_{j}(t,z)$ in the normally ordered form. The
normal ordering formulated in the time-representation allows one to estimate the
means of the Bose-operators over the initial coherent states. As a consequence, the
operator $e^{\hat{O}_{j}(t)}$ takes the form (see \cite{POP99,POP00}):
\begin{equation}\label{teor}
e^{\hat{O}_{j}(t)}=\hat{\mathbf{N}}\exp\left\{\int_{-\infty}^{\infty}
\left[e^{{\mathcal{H}}(\theta)}-1\right]\hat{n}_{0,j}(t-\theta\tau_{r})d\theta\right\},
\end{equation}
where $\theta=t/\tau_{r}$, ${\mathcal{H}}(\theta)=i\gamma\widetilde{h}(\theta)$ and
$\widetilde{h}(\theta)=\tau_{r}h(\theta\tau_r)$. The operators in the integral in
(\ref{teor}) should be understood as the $c$-numbers. Thus, averaging the
$e^{\hat{O}_{j}(t)}$ over the coherent state results
\begin{equation}\label{pp}
\langle e^{\hat{O}_{j}(t)}\rangle=\exp\left\{\int_{-\infty}^{\infty}
\left[e^{{\mathcal{H}}(\theta)}-1\right]\bar{n}_{0,j}(t-\theta\tau_{r})d\theta\right\}
\end{equation}
where
\[\bar{n}_{0,j}(t)=\langle\hat{n}_{0,j}(t)\rangle=\left|\alpha_{0,j}(t)\right|^{2}\]
is the average photon number density of the $j$-th pulse at the input of nonlinear
medium $z=0$. The factorization of the initial coherent states allows one to write
\begin{equation}
\langle\alpha_{0}(t)
|e^{\hat{O}_{1}(t)+\hat{\tilde{O}}_{2}(t)}|\alpha_{0}(t)\rangle=\langle
e^{\hat{O}_{1}(t)}\rangle\langle  e^{\hat{\tilde{O}}_{2}(t)}\rangle,
\end{equation}
where
\begin{eqnarray}
\langle e^{\hat{O}_{1}(t)}\rangle&=&\langle\alpha_{0,1}(t)|
e^{\hat{O}_{1}(t)}|\alpha_{0,1}(t)\rangle,\\ \langle
e^{\hat{\tilde{O}}_{2}(t)}\rangle&=&\langle\alpha_{0,2}(t)|
e^{\hat{\tilde{O}}_{2}(t)}|\alpha_{0,2}(t)\rangle.
\end{eqnarray}

\subsection{Means of Bose-operators and their combinations}

In majority of experimental cases, the parameter $\gamma\ll 1$ and due to this we can
decompose the term in integral in (\ref{pp}) and truncate the decomposition by terms
of the order of $\gamma^{2}$ (see \cite{POP00}). As a result, we get
\begin{eqnarray}
\langle e^{\hat{O}_{1}(t)}\rangle&=&e^{i\phi_{1}(t)-\mu_{1}(t)},\qquad \langle
e^{\hat{O}^{+}_{1}(t)}\rangle=e^{-i\phi_{1}(t)-\mu_{1}(t)},\label{hat1}\\ \langle
e^{\hat{\tilde{O}}_{2}(t)}\rangle
&=&e^{i\tilde{\phi}_{2}(t)-\tilde{\mu}_{2}(t)},\qquad \langle
e^{\hat{\tilde{O}}^{+}_{2}(t)}\rangle=
e^{-i\tilde{\phi}_{2}(t)-\tilde{\mu}_{2}(t)},\label{hat2}
\end{eqnarray}
where
\begin{eqnarray}
\phi_{1}(t)&=&\frac{1}{2}\phi_{0,1}\int_{-\infty}^{\infty}
\widetilde{h}(\theta)r^{2}_{1}(t-\theta\tau_{r})d\theta,\label{ap1}\\
\tilde{\phi}_{2}(t)&=&\frac{1}{2}\tilde{\phi}_{0,2}\int_{-\infty}^{\infty}
\widetilde{h}(\theta)r^{2}_{2}(t-\theta\tau_{r})d\theta,\label{ap2}\\
\mu_{1}(t)&=&\frac{1}{2}\mu_{0,1}\int_{-\infty}^{\infty}
\widetilde{h}^{2}(\theta)r^{2}_{1}(t-\theta\tau_{r})d\theta,\label{ap3}\\
\tilde{\mu}_{2}(t)&=&\frac{1}{2}\tilde{\mu}_{0,2}\int_{-\infty}^{\infty}
\widetilde{h}^{2}(\theta)r^{2}_{2}(t-\theta\tau_{r})d\theta.\label{ap4}
\end{eqnarray}
Here we also introduce the parameters
\begin{eqnarray}
\phi_{0,1}&=&2\gamma_{1}\bar{n}_{0,1}(0),\qquad
\mu_{0,1}=\gamma^{2}_{1}\bar{n}_{0,1}(0)=\gamma_{1}\phi_{0,1}/2,\\
\tilde{\phi}_{0,2}&=&2\tilde{\gamma}\bar{n}_{0,2}(0),\qquad
~\tilde{\mu}_{0,2}=\tilde{\gamma}^{2}\bar{n}_{0,2}(0)=\tilde{\gamma}\phi_{0,2}/2,
\end{eqnarray}
with $r_{j}(t)$ being the envelope of $j$-th pulse so that
$\alpha_{0,j}(t)=\alpha_{0,j}(0)r_{j}(t)$ and $r_{j}(0)=1$. Let us denote for
simplicity $\bar{n}_{0,j}=\bar{n}_{0,j}(0)$. The parameters $\phi_{1}(t)$,
$\mu_{1}(t)$ are certainly connected with SPM of the investigated pulse and the
parameters $\tilde{\phi}_{2}(t)$, $\tilde{\mu}_{2}(t)$ are connected with the XPM of
pulses. The parameters $\phi_{1}(t)$ and $\tilde{\phi}_{2}(t)$ have the physical
meaning of the nonlinear phase additions caused by the SPM and XPM, respectively.
Here $2\gamma_{1}$ represents the nonlinear phase shift per one photon for the
investigated pulse.

A particular interest is connected with the estimation of the average values of
combinations of the exponential Bose-operators over the coherent state. Using the
procedure of normal ordering we have the following formulae:
\begin{eqnarray}
\langle e^{\hat{O}_{1}(t_{1})+\hat{O}_{1}(t_{2})}\rangle &=&
e^{i[\phi_{1}(t_{1})+\phi_{1}(t_{2})]-\mu_{1}(t_{1},t_{2})-{\mathcal{K}}_{1}(t_{1},t_{2})},\label{susi1}\\
\langle e^{\hat{O}^{+}_{1}(t_{1})+\hat{O}_{1}(t_{2})}\rangle &=&
e^{i[-\phi_{1}(t_{1})+\phi_{1}(t_{2})]-\mu_{1}(t_{1},t_{2})+{\mathcal{K}}_{1}(t_{1},t_{2})},\label{susi2}
\end{eqnarray}
and
\begin{eqnarray}
\langle e^{\hat{\tilde{O}}_{2}(t_{1})+\hat{\tilde{O}}_{2}(t_{2})}\rangle &=&
e^{i[\tilde{\phi}_{2}(t_{1})+\tilde{\phi}_{2}(t_{2})]-
\tilde{\mu}_{2}(t_{1},t_{2})-\tilde{\mathcal{K}}_{2}(t_{1},t_{2})},\label{tutu1}\\
\langle e^{\hat{\tilde{O}}^{+}_{2}(t_{1})+\hat{\tilde{O}}_{2}(t_{2})}\rangle &=&
e^{i[-\tilde{\phi}_{2}(t_{1})+\tilde{\phi}_{2}(t_{2})]-
\tilde{\mu}_{2}(t_{1},t_{2})+\tilde{\mathcal{K}}_{2}(t_{1},t_{2})},\label{tutu2}
\end{eqnarray}
where $$\mu_{1}(t_{1},t_{2})=\mu_{1}(t_{1})+\mu_{1}(t_{2}),\qquad
\tilde{\mu}_{2}(t_{1},t_{2})=\tilde{\mu}_{2}(t_{1})+\tilde{\mu}_{2}(t_{2}),$$ and
${\mathcal{K}}_{1}(t_{1},t_{2})$, $\tilde{\mathcal{K}}_{2}(t_{1},t_{2})$ are the time
correlators:
\begin{eqnarray}
{\mathcal{K}}_{1}(t_{1},t_{2})&=&\mu_{0,1}\!
\int_{-\infty}^{\infty}\widetilde{h}(t_{1}-\theta\tau_{r})\widetilde{h}(t_{2}
-\theta\tau_{r})r^{2}_{1}(\theta\tau_{r})d\theta,\label{ss1}\\
\tilde{\mathcal{K}}_{2}(t_{1},t_{2})&=&\tilde{\mu}_{0,2}\!
\int_{-\infty}^{\infty}\widetilde{h}(t_{1}-\theta\tau_{r})\widetilde{h}(t_{2}
-\theta\tau_{r})r^{2}_{2}(\theta\tau_{r})d\theta.\label{ss2}
\end{eqnarray}
As it was mentioned above, our theory makes use of the assumption that
$\tau_{p}\gg\tau_{r}$. Therefore one can simplify expressions
(\ref{ap1})--(\ref{ap4}) and (\ref{ss1}), (\ref{ss2}) eliminating
$r_{j}(t-\theta\tau_{r})$ and $r_{j}(\theta\tau_{r})$ from the integrand in the
particular points: $\theta\tau_{r}=0$ in (\ref{ap1})--(\ref{ap4}), and
$\theta\tau_{r}=t_{1}+\tau/2$ in (\ref{ss1}), (\ref{ss2}), where $\tau=t_{2}-t_{1}$.
Taking into account the response function for the electronic Kerr nonlinearity
(\ref{has}) for the expressions mentioned, we obtain
\begin{eqnarray}
\phi_{1}(t)=\phi_{0,1}\,r^{2}_{1}(t),&\qquad&
\tilde{\phi}_{2}(t)=\tilde{\phi}_{0,2}\,r^{2}_{2}(t),\label{fi1}\\
\mu_{1}(t)=\mu_{0,1}\,r^{2}_{1}(t)/2,&\qquad&
\tilde{\mu}_{2}(t)=\tilde{\mu}_{0,2}\,r^{2}_{2}(t)/2,\label{fi2}
\end{eqnarray}
and the correlators
\begin{eqnarray}
{\mathcal{K}}_{1}(t_{1},t_{2})&=&\mu_{0,1}\,r^{2}_{1}(t_{1}+\tau/2)g(\tau),\label{art1}\\
\tilde{\mathcal{K}}_{2}(t_{1},t_{2})&=&\tilde{\mu}_{0,2}\,r^{2}_{2}(t_{1}+\tau/2)g(\tau).\label{art2}
\end{eqnarray}
The function $g(\tau)$ has the form
\begin{equation}
g(\tau)=\frac{1}{\tau_r}\left(1+\frac{\left|\tau\right|}{\tau_{r}}
\right)\widetilde{h}\left(\frac{\tau}{\tau_{r}}\right).
\end{equation}
The results obtained permit to investigate the statistical properties of the USP in a
medium with the Kerr nonlinearity.

\section{Correlation functions of quadrature components}\label{section4}

Here we restrict our analysis by studying the quadrature components which are defined
by the expressions:
\begin{eqnarray}
\hat{X}_{1}(t,z)&=&\frac{1}{2}\left[\hat{A}_{1}(t,z)+\hat{A}^{+}_{1}(t,z)\right],\\
\hat{Y}_{1}(t,z)&=&\frac{1}{2i}\left[\hat{A}_{1}(t,z)-\hat{A}^{+}_{1}(t,z)\right].
\end{eqnarray}
As it was mentioned above, in the nonlinear media under consideration the photon
statistics of each pulse remains unchanged.

The mean values of the operators $\hat{A}_{1}(t,z)$ and $\hat{A}^{+}_{1}(t,z)$ for
the case of initial pulses being in the coherent state are
\begin{eqnarray}
\langle\hat{A}_{1}(t,z)\rangle&=&\alpha_{0,1}(t)\langle
e^{\hat{O}_{1}(t)}\rangle\langle e^{\hat{\tilde{O}}_{2}(t)}\rangle,\\
\langle\hat{A}^{+}_{1}(t,z)\rangle&=&\alpha^{*}_{0,1}(t)\langle
e^{\hat{O}^{+}_{1}(t)}\rangle\langle e^{\hat{\tilde{O}}^{+}_{2}(t)}\rangle.
\end{eqnarray}
Taking into account Eqs.\ (\ref{hat1}), (\ref{hat2}) and the fact that
$\alpha_{0,j}(t)=|\alpha_{0,j}(t)|e^{i\varphi_{j}(t)}$, for mean values of
quadratures we get
\begin{eqnarray}
\langle\hat{X}_{1}(t,z)\rangle &=& |\alpha_{0,1}\!(t)|e^{-\mu_{1}(t)-
\tilde{\mu}_{2}(t)}\cos{[\Phi_{1}(t)+\tilde{\phi}_{2}(t)]},\label{xip}\\
\langle\hat{Y}_{1}(t,z)\rangle &=& |\alpha_{0,1}\!(t)|e^{-\mu_{1}(t)-
\tilde{\mu}_{2}(t)}\sin{[\Phi_{1}(t)+\tilde{\phi}_{2}(t)]},\label{xix}
\end{eqnarray}
where $\Phi_{1}(t)=\phi_{1}(t)+\varphi_{1}(t)$ and $\varphi_{1}(t)$ is the linear
phase of the investigated pulse. The exponential terms in (\ref{xip}) and (\ref{xix})
are caused by quantum  effects at the SPM and XPM (there are no such terms in the
classical theory). {}From (\ref{xip}) and (\ref{xix}) one can conclude that the
changes of quadratures in time are quasi-statically connected with the changes in
envelopes of both pulses
[$\phi_{1}(t)=\phi_{0,1}r^{2}_{1}(t)=2\gamma_{1}\bar{n}_{0,1}(t)$,
$\tilde{\phi}_{2}(t)=\tilde{\phi}_{0,2}r^{2}_{2}(t)=2\tilde{\gamma}\bar{n}_{0,2}(t)$].
Remind that $t$ is time in the coordinate system moving with the group velocity. In
other words, the causality principle is satisfied for the parameters observed.

We define now the correlation functions of quadrature components
\begin{eqnarray}
R_{X_{1}}(t,t+\tau)&=&\langle\hat{X}_{1}(t,z)\hat{X}_{1}(t+\tau,z)\rangle
-\langle\hat{X}_{1}(t,z)\rangle\langle\hat{X}_{1}(t+\tau,z)\rangle,\\
R_{Y_{1}}(t,t+\tau)&=&\langle\hat{Y}_{1}(t,z)\hat{Y}_{1}(t+\tau,z)\rangle
-\langle\hat{Y}_{1}(t,z)\rangle\langle\hat{Y}_{1}(t+\tau,z)\rangle.
\end{eqnarray}
The analysis of the correlation functions requires the evaluation of the correlators
\[C_{X_{1}}(t_{1},t_{2})=\langle\hat{X}_{1}(t_{1})\hat{X}_{1}(t_{2})\rangle \qquad
\mbox{and}\qquad
C_{Y_{1}}(t_{1},t_{2})=\langle\hat{Y}_{1}(t_{1})\hat{Y}_{1}(t_{2})\rangle.\] Using
the permutation relations (\ref{permut1}), (\ref{permut2}) and Eqs.\
(\ref{susi1})--(\ref{tutu2}) we obtain
\begin{eqnarray}
C_{X_{1}}(t_{1},t_{2})&=&\frac{1}{4}\delta(t_{2}-t_{1})
+\frac{1}{2}\left|\alpha_{0,1}(t_{1})\right|
\left|\alpha_{0,1}(t_{2})\right|e^{-\tilde{\mu}_{1,2}(t_{1},t_{2})}\nonumber\\
&{}&\times\Bigl\{e^{-\tilde{\Lambda}_{1,2}(t_{1},t_{2})}
\cos{[\tilde{\Phi}_{1,2}(t_{1})+\tilde{\Phi}_{1,2}(t_{2})+
\gamma_{1}\widetilde{h}(t_{2}-t_{1})]}\nonumber\\
&{}&+e^{\tilde{\Lambda}_{1,2}(t_{1},t_{2})}\cos{[\tilde{\Phi}_{1,2}(t_{1})
-\tilde{\Phi}_{1,2}(t_{2})]}\Bigl\},\\
C_{Y_{1}}(t_{1},t_{2})&=&\frac{1}{4}\delta(t_{2}-t_{1})
-\frac{1}{2}\left|\alpha_{0,1}(t_{1})\right|
\left|\alpha_{0,1}(t_{2})\right|e^{-\tilde{\mu}_{1,2}(t_{1},t_{2})}\nonumber\\
&{}&\times\Bigl\{e^{-\tilde{\Lambda}_{1,2}(t_{1},t_{2})}\cos{[\tilde{\Phi}_{1,2}(t_{1})+
\tilde{\Phi}_{1,2}(t_{2})+\gamma_{1}\widetilde{h}(t_{2}-t_{1})]}\nonumber\\
&{}&-e^{\tilde{\Lambda}_{1,2}(t_{1},t_{2})}\cos{[\tilde{\Phi}_{1,2}(t_{1})-\tilde{\Phi}_{1,2}(t_{2})]}\Bigl\},
\end{eqnarray}
where we introduced for simplicity the notations
\begin{equation}
\eqalign{\tilde{\mu}_{1,2}(t_1,t_2)&=\mu_{1}(t_1,t_2)+\tilde{\mu}_{2}(t_1,t_2),\\
\tilde{\Lambda}_{1,2}(t_{1},t_{2})&=\Lambda_{1}(t_1,t_2)+\tilde{\Lambda}_{2}(t_{1},t_{2}),\\
\Lambda_{1}(t_{1},t_{2})&=\mu_{1}(t_1,t_2)\widetilde{h}(t_2-t_1),\\
\tilde{\Lambda}_{2}(t_{1},t_{2})&=\tilde{\mu}_{2}(t_{1},t_{2})\widetilde{h}(t_{2}-t_{1}),\\
\tilde{\Phi}_{1,2}(t)&=\Phi_{1}(t)+\tilde{\phi}_{2}(t).}
\end{equation}
As a result, for the correlation functions of quadrature components we finally get:
\begin{eqnarray}
R_{X_{1}}(t,t+\tau)&=&\frac{1}{4}\Bigl\{\delta(\tau)
-\phi_{1}(t)h(\tau)\sin{2\tilde{\Phi}_{1,2}(t)}\nonumber\\
&{}&+\bigl[\phi^{2}_{1}(t)+\tilde{\phi}_{1}(t)\tilde{\phi}_{2}(t)\bigl]
g(\tau)\sin^{2}{\tilde{\Phi}_{1,2}(t)}\Bigl\},\label{wu1}\\
R_{Y_{1}}(t,t+\tau)&=&\frac{1}{4}\Bigl\{\delta(\tau)
+\phi_{1}(t)h(\tau)\sin{2\tilde{\Phi}_{1,2}(t)}\nonumber\\
&{}&+\bigl[\phi^{2}_{1}(t)+\tilde{\phi}_{1}(t)\tilde{\phi}_{2}(t)\bigl]
g(\tau)\cos^{2}{\tilde{\Phi}_{1,2}(t)}\Bigl\}.\label{wu2}
\end{eqnarray}
To obtain (\ref{wu1}) and (\ref{wu2}), the $\gamma\ll 1$ and $\tau_{r}\ll\tau_{p}$
approximations have been used.


\section{Spectrum of quantum fluctuations of quadrature components}\label{section5}

The spectral densities of quantum fluctuations of the quadrature components are
defined by the expression:
\begin{equation}
S_{X_{1},Y_{1}}(\omega,t)=\int_{-\infty}^{\infty}
R_{X_{1},Y_{1}}(t,t+\tau)e^{i\omega\tau}d\tau.
\end{equation}
Taking into account a slow change of the envelope during the relaxation time, one
arrives at:
\begin{eqnarray}
\fl S_{X_{1}}(\omega,t)&=&\!\frac{1}{4}\Bigl\{1-2\phi_{1}(t)L(\omega)
\sin{2\tilde{\Phi}_{1,2}(t)}+4[\phi^{2}_{1}(t)\smash{+}\tilde{\phi}_{1}(t)\tilde{\phi}_{2}(t)]L^{2}(\omega)\sin^{2}{\tilde{\Phi}_{1,2}(t)}\Bigl\},\label{sio}\\
\fl
S_{Y_{1}}(\omega,t)&=&\!\frac{1}{4}\Bigl\{1+2\phi_{1}(t)L(\omega)\sin{2\tilde{\Phi}_{1,2}(t)}
+4[\phi^{2}_{1}(t)\smash{+}\tilde{\phi}_{1}(t)\tilde{\phi}_{2}(t)]L^{2}(\omega)\cos^{2}{\tilde{\Phi}_{1,2}(t)}\Bigl\},\label{ert}
\end{eqnarray}
where $L(\omega)=[1+(\omega\tau_{r})^{2}]^{-1}$ and
$\tilde{\phi}_{j}(t)=2\tilde{\gamma}\bar{n}_{0,j}(t)$.

From (\ref{sio}) and (\ref{ert}) it follows that in the case $\tilde{\phi}_{2}(t)=0$,
i.e. in the absence of XPM effect, the known result at the SPM effect for the
investigated pulse can be obtained \cite{POP99,POP00}. At the SPM process, one can
control the spectrum by choosing the phase of the initial investigated light pulse to
be optimal for the chosen frequency \cite{POP00,POP0}. The presence of the XPM effect
adds new terms in the multiplier and the phase in  expressions (\ref{sio}) and
(\ref{ert}) in comparison with the ones for the SPM effect only. This circumstance
gives us another possibility to control the fluctuation spectrum of the investigated
pulse by varying these terms.

{}From (\ref{sio}) and (\ref{ert}) it also follows that the choice of the phase
$\tilde{\Phi}_{1,2}(t)$ determines the level of quantum fluctuations to be lower or
higher than the shot-noise level $S_{X_{1}}(\omega)=S_{Y_{1}}(\omega)=1/4$,
corresponding to the coherent state of the initial pulse. In agreement with the
Heisenberg uncertainty relation, the spectrum of the $Y_{1}$-quadrature is shifted by
a phase $\pi/2$ in comparison with the spectrum of the $X_{1}$-quadrature.

If one chooses the optimal phase of the initial investigated  pulse
\begin{equation}\label{phase1}
\varphi_{0,1}(t)=\frac{1}{2}\arctan{\left(\frac{\phi_{1}(t)}
{\phi^{*}_{1,2}(t)L(\Omega_{0})}\right)}-\phi_{1}(t)-\tilde{\phi}_{2}(t)
\end{equation}
for the reduced frequency $\Omega_{0}=\omega_{0}\tau_{r}$, then spectral densities
(\ref{sio}), (\ref{ert}) take the form
\begin{eqnarray}
\fl S^{0}_{X_{1}}(\Omega_{0},t)&=&\frac{1}{4}\Bigl\{1-2L(\Omega_{0})
\bigl[\phi^{2}_{1}(t)+\phi^{{*}^{2}}_{1,2}(t)L^{2}(\Omega_{0})\bigl]^{1/2}
+2\phi^{*}_{1,2}(t)L(\Omega_{0})\Bigl\},\label{So}\\ \fl
S^{0}_{Y_{1}}(\Omega_{0},t)&=&\frac{1}{4}\Bigl\{1+2L(\Omega_{0})\bigl[\phi^{2}_{1}(t)+
\phi^{{*}^{2}}_{1,2}(t)L^{2}(\Omega_{0})\bigl]^{1/2}
+2\phi^{*}_{1,2}(t)L(\Omega_{0})\Bigl\},\label{Soe}
\end{eqnarray}
where $\phi^{{*}}_{1,2}(t)=\phi^{2}_{1}(t)+\tilde{\phi}_{1}(t)\tilde{\phi}_{2}(t)$.

At any frequency $\Omega$, we have
\begin{eqnarray}
\fl S_{X_{1}}(\Omega,t)&=&S^{0}_{X_{1}}(\Omega_{0},t)+\frac{1}{2}
[L(\Omega)-L(\Omega_{0})]\Bigl\{\phi^{*}_{1,2}(t)[L(\Omega)+L(\Omega_{0})]\nonumber\\
\fl &{}&-\bigl\{\phi^{2}_{1}(t)+\phi^{{*}^{2}}_{1,2}(t)L(\Omega_{0})
[L(\Omega)+L(\Omega_{0})]\bigl\}
\bigl[\phi^{2}_{1}(t)+\phi^{{*}^{2}}_{1,2}(t)L^{2}(\Omega_{0})
\bigl]^{-1/2}\Bigl\},\label{sist}\\ \fl
S_{Y_{1}}\!(\Omega,t)&=&S^{0}_{Y_{1}}(\Omega_{0},t)+\frac{1}{2}
[L(\Omega)-L(\Omega_{0})]\Bigl\{\phi^{*}_{1,2}(t)[L(\Omega)+L(\Omega_{0})]\nonumber\\
\fl &{}&+\bigl\{\phi^{2}_{1}(t)+\phi^{{*}^{2}}_{1,2}(t)L(\Omega_{0})
[L(\Omega)+L(\Omega_{0})]\bigl\}
\bigl[\phi^{2}_{1}(t)+\phi^{{*}^{2}}_{1,2}(t)L^{2}(\Omega_{0})
\bigl]^{-1/2}\Bigl\}.\label{sios}
\end{eqnarray}

{}From (\ref{sist}) and (\ref{sios}) it follows that the change
$\tilde{\phi}_{2}(t)=2\tilde{\gamma}\bar{n}_{0,2}(t)$ provides a possibility for
controlling the spectrum of quadrature-component fluctuations. Certainly, this change
can be realized by varying the control pulse intensity $\bar{n}_{0,2}(t)$ at the
optimal phase of the investigated pulse.

\begin{figure}
\centering
\includegraphics[height=.5\textwidth]{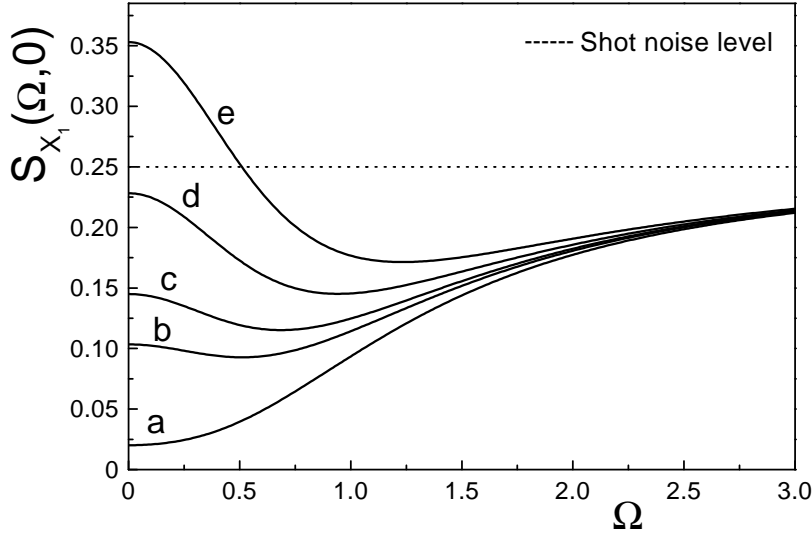}
\caption{Fluctuation spectrum of the quadrature-squeezed component of the
investigated pulse for the case of optimal initial phase $\varphi_{0,1}(t)$ chosen
for $\Omega_{0}=0$ and for different intensities of the control pulse
$\bar{n}_{0,2}/\bar{n}_{0,1}=0$ (a), $2$ (b), $3$ (c), $5$ (d), $8$ (e). Curves are
calculated for the time moment $t=0$ and $\gamma_1=\gamma_2=2\tilde{\gamma}$,
$\phi_{0,1}=2\gamma_{1}\bar{n}_{0,1}=2$.} \label{fig1}
\end{figure}

\begin{figure}
\centering
\includegraphics[height=.5\textwidth]{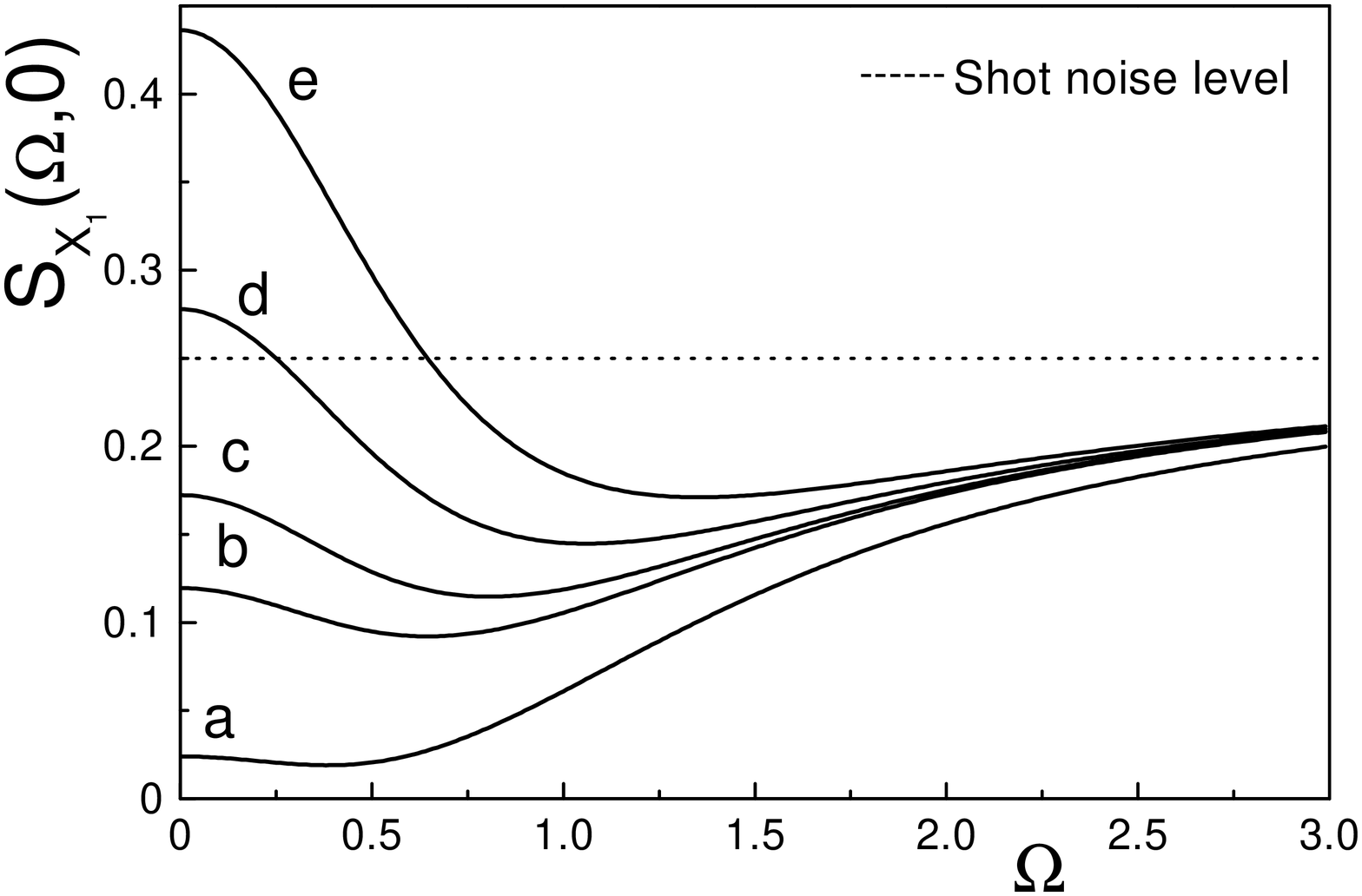}
\caption{The same as in Fig.\ \ref{fig1}\ but for $\Omega_{0}=0.5$.} \label{fig2}
\end{figure}

\begin{figure}
\centering
\includegraphics[height=.5\textwidth]{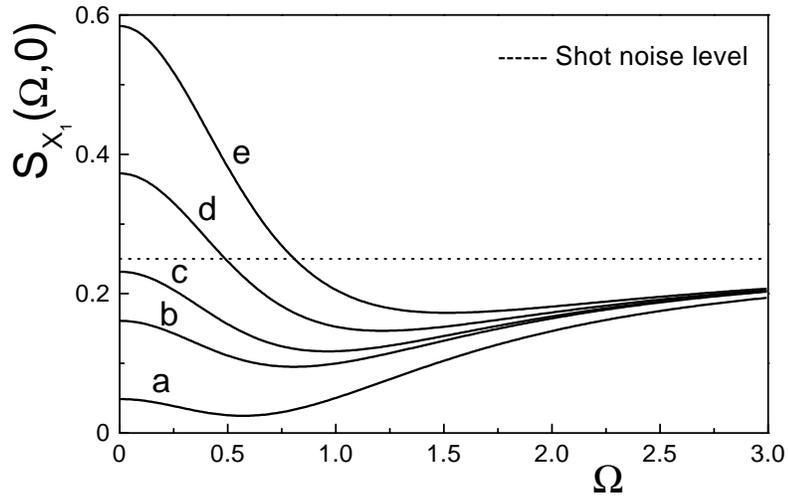}
\caption{The same as in Fig.\ \ref{fig1}\ but for $\Omega_{0}=0.7$.}  \label{fig3}
\end{figure}

The spectra of the investigated pulse at a fixed optimal initial phase
$\varphi_{0,1}(t)$ (chosen at $\Omega_{0}=0$, $0.5$, $0.7$ and the time moment $t=0$)
for different values of the photon numbers of the control pulse are displayed in
Figs.\ \ref{fig1}, \ref{fig2}, and \ref{fig3}, respectively.

{}From Figs.\ \ref{fig1}-\ref{fig3}\ it follows that the change of the intensity of
one light pulse gives a method for the control of squeezed spectra formation for
another pulse and this control is effective if the intensity of the control pulse is
much larger than that of the investigated pulse. While the intensity of the control
pulse increases, the squeezing at the frequency for which the initial phase was
chosen to be optimal, is destroyed. The suppression of the quadrature-component
fluctuations for the investigated light pulse at higher intensities of the control
pulse takes place in the frequency domain $\omega\approx1/\tau_{r}$.

\begin{figure}
\centering
\includegraphics[height=.5\textwidth]{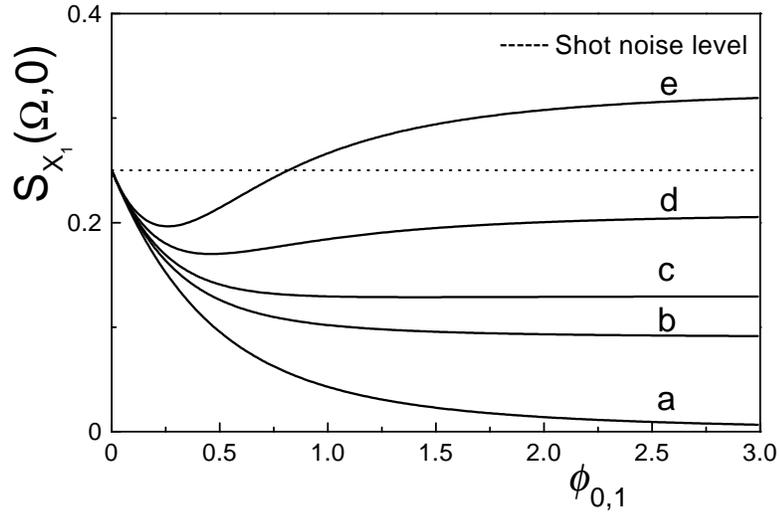}
\caption{Fluctuation spectrum of the quadrature-squeezed component  of the
investigated pulse for the frequency $\Omega=0$ for the case of optimal initial phase
$\varphi_{0,1}(t)$ chosen for $\Omega_{0}=0$ and for different intensities of the
control pulse $\bar{n}_{0,2}/\bar{n}_{0,1}=0$ (a), $2$ (b), $3$ (c), $5$ (d), $8$
(e). Curves are calculated for the time moment $t=0$ and
$\gamma_1=\gamma_2=2\tilde{\gamma}$.} \label{fig4}
\end{figure}

\begin{figure}
\centering
\includegraphics[height=.5\textwidth]{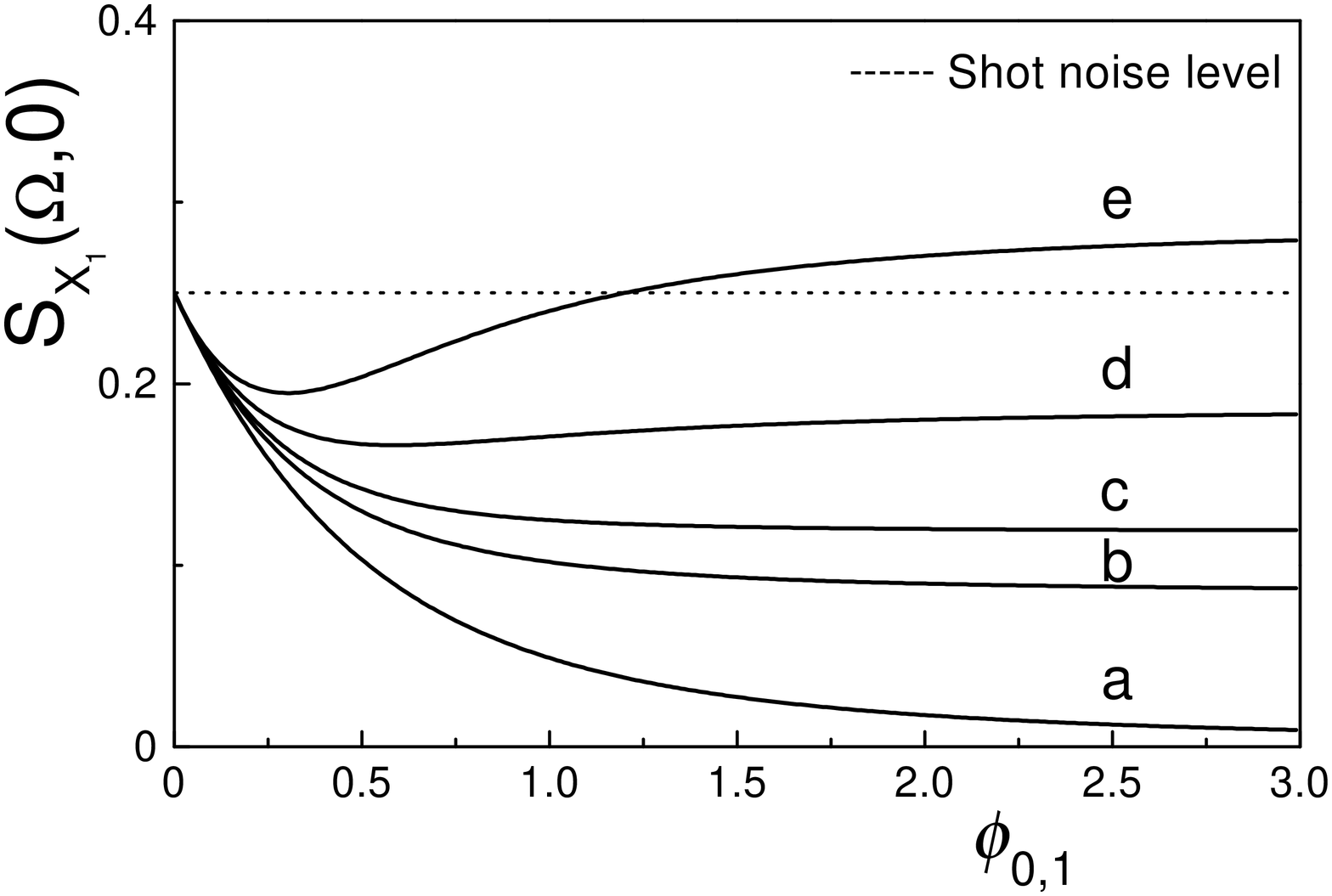}
\caption{The same as in Fig.\ \ref{fig4}\ but for the frequency $\Omega=0.3$.}
\label{fig5}
\end{figure}

\begin{figure}
\centering
\includegraphics[height=.5\textwidth]{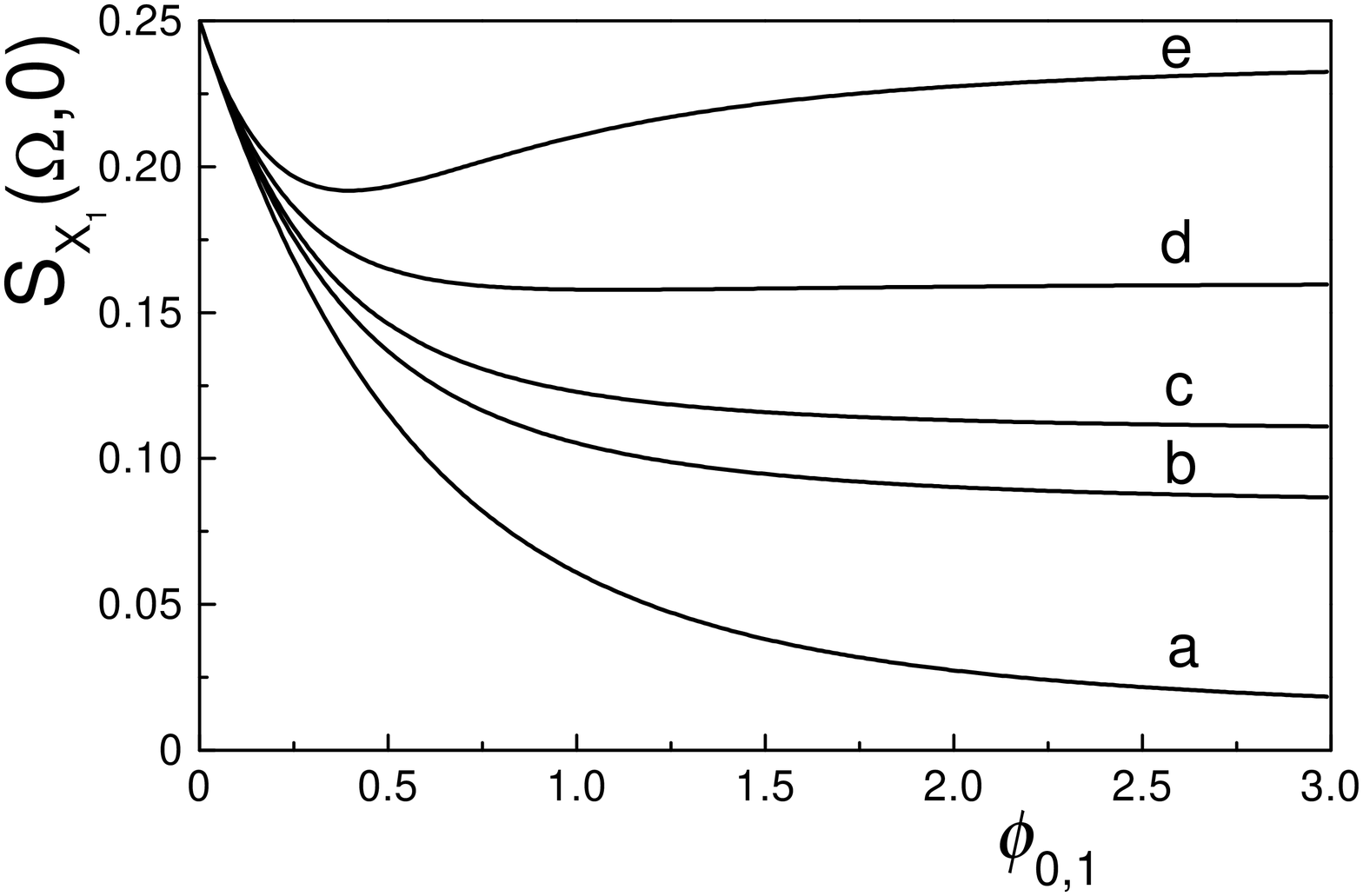}
\caption{The same as in Fig.\ \ref{fig4}\ but for the frequency $\Omega=0.5$.} \label{fig6}
\end{figure}

Squeezing spectra of the investigated pulse (\ref{sist}) for various frequencies
$(\Omega=0$, $0.3$, $0.5)$ and different intensities of the control pulse as
functions of the maximum nonlinear phase addition $\phi_{0,1}$ are shown in Figs.\
\ref{fig4}, \ref{fig5}, and \ref{fig6}, respectively. One can see from Figs.\
\ref{fig4}-\ref{fig6} that at higher intensities of the control pulse the spectral
density under investigation reaches the minimum value in the domain $\phi_{0,1}\leq
1$ and then the spectral density does not practically change.

\begin{figure}
\centering
\includegraphics[height=.5\textwidth]{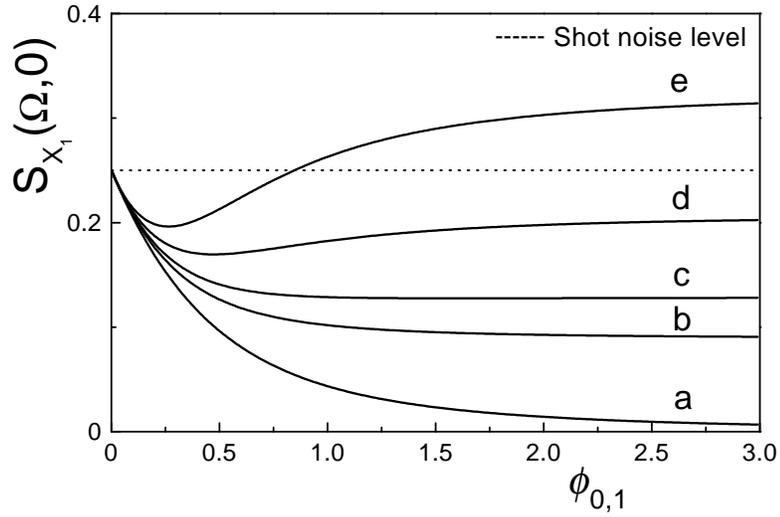}
\caption{Fluctuation spectrum of the quadrature-squeezed component of investigated
pulse at the frequency $\Omega=0.5$ for the case of optimal initial phase
$\varphi_{0,1}(t)$ chosen for $\Omega_{0}=0.5$ and for different intensities of the
control pulse $\bar{n}_{0,2}/\bar{n}_{0,1}=0$ (a), $2$ (b), $3$ (c), $5$ (d), $8$
(e). Curves are calculated for the time moment $t=0$ and
$\gamma_1=\gamma_2=2\tilde{\gamma}$.} \label{fig7}
\end{figure}

The appropriate for our purpose spectra can be also obtained by changing the
intensity of control pulse and measuring the spectra for defined frequency $\Omega$
when the initial phase of the investigated pulse is chosen to be optimal for various
$\Omega_{0}$. Such a dependence is displayed in Fig.\ \ref{fig7}. In some sense, the
spectra shown in Figs.\ \ref{fig4}\ and \ref{fig7}\ are similar, and they demonstrate
that the measurements carried out at frequencies, for which the initial phase was
chosen optimal, give almost the same results. {}From  Figs.\ \ref{fig4}\--\ref{fig7}\
one can see that the XPM effect determines the domain of $\phi_{0,1}$ values where
the spectral density of quadrature component fluctuations is independent on the
intensity of the investigated pulse at the output of nonlinear medium.

\section{Conclusion}\label{conclusions}

We have developed the quantum theory of two USPs propagation in the nonlinear medium
with inertial electronic Kerr nonlinearity. In the considered case, the propagating
pulses are subject simultaneously to the SPM and XPM effects. The nonlinear medium is
suggested to be lossless and dispersionless, that is, the pulse frequencies are
off-resonances. Nevertheless, to develop in this case the consistent quantum theory
of the SPM--XPM effect, it is necessary to take into account a finite time of
nonlinear response. In the developed approach, we  neglect thermal fluctuations of
nonlinearity, and the causality principle is not violated for the observed values.

The nonlinear response time defines spectral width of quadrature-squeezed light [see
Eqs.\ (\ref{sio}), (\ref{ert})]. The level of suppression of quantum fluctuations
depends on the nonlinear phase additions due to both the SPM and XPM phenomena.

It is shown that the frequency, at which the suppressions of fluctuations is maximum,
can be controlled by adjusting the initial phase of the investigated light pulse and
the intensity of another pulse.

The results of this paper can be used in the study of creation of the USPs in
nonclassical states and in quantum non-demolition measurements using USPs.

The approach developed can be applied to generalize the theory of other quantum
effects associated with the two-mode light propagation in the Kerr media (see
references in \cite{Tan2}) to the case of pulse fields, for example, generation of
polarization-squeezed light \cite{Orlov}. This study is in progress.

\ack{The work was partially supported by the Russian Foundation for Basic Research
under Project No.~01-02-16311 and by the Ministry for Industry, Science and
Technology of the Russian Federation.}

\end{document}